\documentclass[twocolumn,aps,prb,raggedbottom,nobalancelastpage,amssymb,superscriptaddress]{revtex4-1}

\usepackage{amsmath}
\usepackage{amssymb}
\usepackage{amsfonts}
\usepackage{dsfont}
\usepackage{graphicx}
\usepackage{bm}
\usepackage{color,soul}
\usepackage{appendix}
\usepackage{siunitx}

\def \Fst{1^\textrm{st}}
\def \Snd{2^\textrm{nd}}

\def \l{\lambda}

\renewcommand{\figurename}{Figure}

\begin{document}
\newcommand{\comm}[1]{\hl{#1}}
\newcommand{\commb}[1]{{\sethlcolor{cyan}\hl{#1}}}

\makeatletter
\renewcommand*{\@fnsymbol}[1]{%
  \ensuremath{\dagger}
}
\makeatother

\title{Photonic topological pumping through the edges of a dynamical four-dimensional quantum Hall system}

\author{Oded Zilberberg}
\affiliation{Institute for Theoretical Physics, ETH Zurich, 8093 Z{\"u}rich, Switzerland}

\author{Sheng Huang}
\affiliation{\mbox{Department of Electrical and Computer Engineering, University of Pittsburgh, Pittsburgh, Pennsylvania 15261, USA}}

\author{Jonathan Guglielmon}
\affiliation{Department of Physics, The Pennsylvania State University, University Park, Pennsylvania 16802, USA}

\author{Mohan~Wang}
\affiliation{\mbox{Department of Electrical and Computer Engineering, University of Pittsburgh, Pittsburgh, Pennsylvania 15261, USA}}

\author{Kevin Chen}
\affiliation{\mbox{Department of Electrical and Computer Engineering, University of Pittsburgh, Pittsburgh, Pennsylvania 15261, USA}}

\author{Yaacov E. Kraus}
\thanks{Deceased 9 November 2016.}
\affiliation{Department of Physics, Holon Institute of Technology, Holon 5810201, Israel}

\author{Mikael C. Rechtsman}
\affiliation{Department of Physics, The Pennsylvania State University, University Park, Pennsylvania 16802, USA}

\maketitle

\textbf{When a two-dimensional electron gas is exposed to a perpendicular magnetic field and an in-plane electric field, its conductance becomes {\it quantized} in the transverse in-plane direction: this is known as the quantum Hall (QH) effect \cite{klitzing1980new}. This effect is a result of the nontrivial topology of the system's electronic band structure, where an integer topological invariant known as the $\Fst$ Chern number leads to the quantization of the Hall conductance \cite{thouless1982quantized}. Interestingly, it was shown that the QH effect can be generalized mathematically to four spatial dimensions (4D)~\cite{avron1989,IQHE4D}, but this effect has never been realized for the obvious reason that experimental systems are bound to three spatial dimensions.  In this work, we harness the high tunability and control offered by photonic waveguide arrays to experimentally realize a dynamically-generated 4D QH system using a 2D array of coupled optical waveguides. The inter-waveguide separation is constructed such that the propagation of light along the device samples over higher-dimensional momenta in the directions orthogonal to the two physical dimensions, thus realizing a 2D topological pump~\cite{LaughlinNobelLecture,thouless1982quantized,Zilberberg:2012a,Zilberberg:2013b,Zilberberg:2015a,Zilberberg:2016b,nakajima2016}. As a result, the device's band structure is associated with 4D topological invariants known as $\Snd$ Chern numbers which support a quantized bulk Hall response with a 4D symmetry~\cite{Zilberberg:2013b}. In a finite-sized system, the 4D topological bulk response is carried by localized edges modes that cross the sample as a function of of the modulated auxiliary momenta. We directly observe this crossing through photon pumping from edge-to-edge and corner-to-corner of our system.  These are equivalent to the pumping of charge across a 4D system from one 3D hypersurface to the opposite one and from one 2D hyperedge to another, and serve as first experimental realization of higher-dimensional topological physics.}

The mathematical field of topology manifests naturally in solid state systems~\cite{RMP_TI,RMP_TI2}. In insulators, electrons populate the electronic states below the spectral gap of the system. These states can be mathematically mapped onto abstract shapes depending on 
their geometric properties and can therefore be characterized by a topological invariant~\cite{RMP_TI,RMP_TI2}.  The realization that these topological invariants manifest as quantized bulk responses, as well as through corresponding topologically protected boundary states, has revolutionized the way we understand material properties. A wide variety of fields have explored these topological phenomena beyond solid-state materials, including in photonic \cite{haldane2008possible, wang2009observation, rechtsman2012, Hafezi:2013}, ultracold atomic \cite{dalibard2011colloquium, aidelsburger2013realization, jotzu2014experimental}, and phononic \cite{kane2014topological,huber} systems. 

The introduction of topological concepts into photonics in particular~\cite{haldane2008possible} has opened up many exciting avenues of research. Much of this activity has been focused on the experimental observation of topologically-protected edge states in systems ranging from photonic crystals and metamaterials in the microwave domain~\cite{wang2009observation,KhanikaevNatMat1, KhanikaevNatMat2}, to arrays of coupled waveguides~\cite{rechtsman2012, Zilberberg:2012a} and integrated silicon ring resonators in the visible domain~\cite{Hafezi:2013}. In all of these works, spatially-periodic dielectric structures act as lattices for light which, in combination with an engineered synthetic gauge field, lead to topological 2D photonic energy bands. Going beyond 2D, the first experimental works on 3D lattices have recently unveiled intriguing topological features in their photonic band structures \cite{Khanikaev:2017} such as Weyl points~\cite{Lu:2013,Lu:2015, Rechtsman:2017}.  

Systems such as these fit well into the study of our 3D world and its constituents. The study of topological phases can, nevertheless, be defined and understood mathematically in higher dimensions, with a hallmark example being the 4D generalization of the 2D quantum Hall effect~\cite{avron1989,IQHE4D, Zilberberg:2013b}. In 2D QH systems, energy bands of the electronic band structure are characterized by the $\Fst$ Chern number~\cite{TKNN,Avron1}, which quantizes the Hall conductance and thus counts the number of 1D chiral edge states in a finite system. In 4D, energy bands can be characterized by another topological invariant - the $\Snd$ Chern number~\cite{avron1989,IQHE4D, QiZhang, Zilberberg:2013b}. Similarly to the 2D case, the 4D invariant manifests through an additional quantized bulk response that has corresponding 4D hypersurface phenomena. Until recently, the latter seemed of pure theoretical interest, simply because its realization would have required four spatial dimensions. The control and flexibility of atomic and photonic systems, however, have inspired recent proposals to include synthetic dimensions in the attempt of realizing higher-dimensional topological systems directly \cite{MaciejSynthetic, HrvojeSynthetic,Zilberberg:2015b,Zilberberg:2016a}. Accessing higher-dimensional systems, therefore, poses a realistic new frontier for studying fundamental physics.

The concept of `topological pumps' lends itself well to the incorporation of synthetic dimensions and the study of higher-dimensional physics. For example, one may consider a family of 1D systems parametrized by a momentum in a fictitious orthogonal dimension. This momentum is known as the `pump parameter' and it acts as an auxiliary dimension, thus mapping the 1D pump to the 2D QH system and to its characterization by a $\Fst$ Chern number \cite{thouless1982quantized, Zilberberg:2012a}. The topological bulk response of the 1D pump matches that of the 2D QH effect: varying the pump parameter effectively generates an electro-motive force that pushes charge across the physical dimension, where an integer number of charges is pumped per cycle in accordance with the $\Fst$ Chern number~\cite{LaughlinNobelLecture,thouless1982quantized}. Recently, the quantized bulk response of 1D topological pumps has been demonstrated in cold atom experiments~\cite{Zilberberg:2016b,nakajima2016} and its corresponding boundary states were addressed in photonic coupled-waveguide arrays~\cite{Zilberberg:2012a,Zilberberg:2015a}. 

\begin{figure}
\includegraphics[width=\columnwidth]{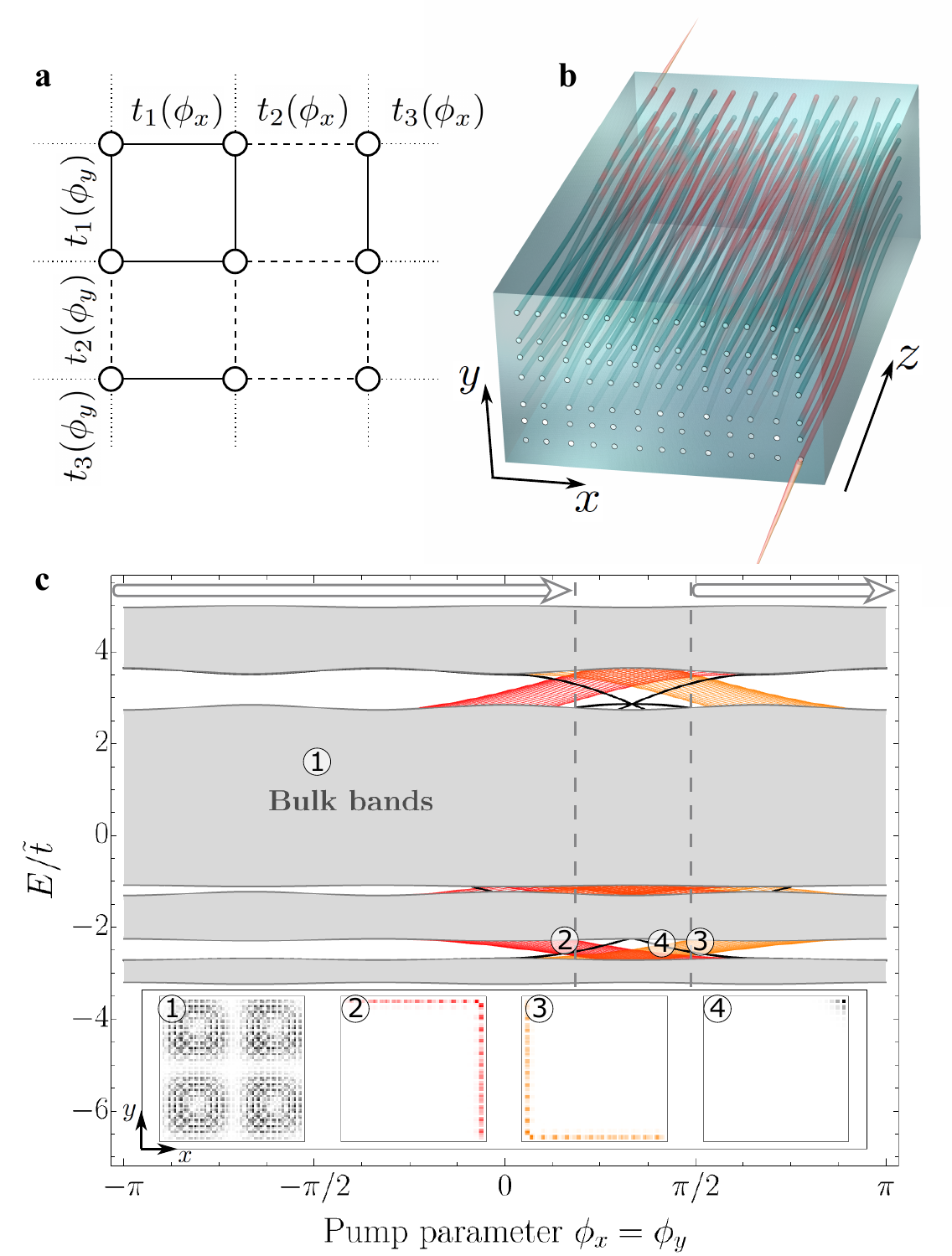}
\caption{\textbf{The two-dimensional topological pump and its corresponding band structure.} \textbf{a,} A schematic diagram of the lattice model [Eq.~\eqref{Eq:H2D}] with a $3\times 3$ unit cell ($b_x=1/3, b_y=1/3$) having three different hopping amplitudes, in each direction (solid, dashed, dotted lines), that can be modulated using the pump parameters $\phi_x$ and $\phi_y$. We assume here periodic boundary conditions. \textbf{b,} An illustration of the 2D array of evanescently-coupled waveguides used in the experiment with $z$-dependent waveguide spacings and $7\times 13$ dimensions. Light is injected into the input facet of the device, it pumps across it during its propagation (due to the topological nature of the 2D pump), and is collected on the other side using a CCD camera. \textbf{c,} Calculated band structure for a similar device consisting of a $70\times 70$ array of coupled waveguides taken along the path $ \phi_x = \phi_y$ (the larger dimensions are chosen for clarity of presentation), at wavelength $1550nm$. Bulk modes are shown in gray, edge modes in red and orange, and corner modes in black. Insets show representative wavefunctions for each type of mode. 
Due to the long-range hopping in the device, when the various edge-modes of the 2D pump are degenerate they can hybridize to form a right-angle wedge. Similarly, the corner modes vanish into the bulk bands along their pump-path and generally hybridize with bulk modes. We perform pumping experiments to study the properties of these topological boundary states, where $\phi_i$ are scanned between $0.477\pi$ and $2.19\pi$ in pump-parameter space (marked by vertical dashed lines and arrow marking the direction of the pumping), see Figs.~\ref{fig2} and \ref{fig3}. }\label{fig1}
\end{figure}

Interestingly, a 2D topological pump can be subject to two pump parameters such that it corresponds to a 4D QH system~\cite{Zilberberg:2013b}. In its simplest form, the 4D QH system can be understood as a direct sum of two 2D QH systems in disjoint planes~\cite{Zilberberg:2013b,Zilberberg:2015b,Zilberberg:2016a}. Correspondingly, the 2D topological pump can manifest as a direct sum of two 1D pumps in orthogonal axes~\cite{Zilberberg:2013b}. Here, we consider `off-diagonal' pumps where the hopping amplitudes are modulated as a function of pump parameters \cite{Zilberberg:2012a,Zilberberg:2015a}, i.e., we study a 2D tight-binding model of particles that hop on a lattice described by the Hamiltonian (see Fig.~\ref{fig1}\textbf{a})
\begin{align} \label{Eq:H2D}
    &H = \sum_{x,y} t_x(\phi_x) c_{x,y}^\dag c_{x+1,y}+ t_y(\phi_y) c_{x,y}^\dag c_{x,y+1} + \text{h.c.}\,,
\end{align}
where $c_{x,y}$ is the annihilation operator of a particle at site $(x,y)$; $t_i(\phi_i)=\tilde{t}_i + \l_i \cos( 2\pi b_i i + \phi_i)$ are modulated hopping amplitudes in the $i=x,y$ directions with bare hopping $\tilde{t}_i$ and modulation amplitudes $\l_i$. The modulation frequencies $b_i$ can be mapped in 4D to two magnetic fields threading the $x-v$ and $y-w$ planes. The pump parameters $\phi_x$ and $\phi_y$ correspond to momenta in the $v-$ and $w-$directions, i.e., they can be understood as a dynamically generated electric field perturbations in these directions, respectively. Considering that the pump parameters correspond to additional effective dimensions, we can characterize spectral bands of the 2D pump with nontrivial $\Snd$ Chern numbers that manifest in a quantized bulk response with a 4D symmetry~\cite{Zilberberg:2013b}.

In this work, we realize such a 2D topological pump using photonic coupled waveguide arrays (see Figs.~\ref{fig1}\textbf{b}). Each waveguide array is constructed to emulate the 2D pump model [Eq.~\eqref{Eq:H2D}] with $b_x=1/3, b_y=1/3$, and with 7 rows and 13 columns. The inter-waveguide separation is taken such that the evanescent coupling between nearest-neighboring waveguides is modulated according to the hopping amplitudes of Eq.~\eqref{Eq:H2D}, with $\lambda_x=\lambda_y=1.06/cm$ and $\tilde{t}_x=\tilde{t}_y=1.94/cm$ (at $1550nm$ wavelength). Nevertheless, the evanescent coupling is a function of both separation and wavelength (see Methods). Therefore, the resulting structure has coupling between waveguides beyond its nearest neighbors and the emulated model is not a pure direct sum of two disjoint 1D pumps. Despite this deformation, the calculated spectrum for the device manifests bulk gaps and bands with gap-traversing boundary states showing appearances of both edge and corner states - see Fig.~\ref{fig1}\textbf{c} and its insets, and see Supplementary Information (including Supplementary Information Fig.~\ref{figSupp}) for a comparison with the direct sum spectrum.  

One can understand the appearance of such topological edge phenomena to be a manifestation of the nontrivial 4D topology of the 2D pump. Let us first study the structure of the model described in Eq.~\eqref{Eq:H2D}.  Since this model can be decomposed as a direct sum of 1D pumps that each have gaps that are associated with nontrivial $\Fst$ Chern numbers~\cite{TKNN,Avron1,Zilberberg:2012a}, it follows that: (a) the spectrum of the 2D pump can be understood as a Minkowski sum of the two independent 1D pump spectra, $E=E_x + E_y$, (b) the states of the model are product states of the two independent models, and (c) the product bands are associated with $\Snd$ Chern numbers equal to the multiplication of the individual $\Fst$ Chern numbers. The latter manifests in nontrivial bulk phenomena only when gaps remain open in the resulting Minkowski sum spectrum. Importantly, such $\Snd$ Chern number response and its 4D symmetry imply that charge will be pumped as a response to a scan of either of the pump parameters, $\phi_i$ as well as to both, see Supplementary Information for more detail.

Let us now consider these properties of the model \eqref{Eq:H2D} in an open geometry.  As each topological 1D pump has 1D bulk modes as well as 0D boundary modes, (a) and (b) imply that we can group the 2D pump states into three categories: (i) 2D bulk modes composed of products of 1D bulk modes, (ii) edge modes composed of products of 1D bulk modes with a 0D boundary, and (iii) corner modes that are a product of 0D boundaries. 
The boundary modes [cases (ii) and (iii)], are those that support the quantized $\Snd$ Chern number response in an finite-sized system (see Supplementary Information for details). Interestingly, the 1D edge states of the 2D physical system maps onto 3D hypersurface states in the 4D picture and 0D corner states map to protected 2D hyperplanes, thus, illuminating the intricate 4D hypersurface phenomena associated with the $\Snd$ Chern number.

Our coupled-waveguide device is, however, not a simple direct sum of two 1D pumps in disjoint axes due to longer-ranged hopping. Nevertheless, the perturbation induced by the long-range hopping keeps the bulk gaps open. As result, the topological characterization of these gaps by nontrivial $\Snd$ Chern numbers implies that the bulk response must remain unchanged.  The appearance of edge states that traverse the gaps as a function of the pump-parameters $\phi_i$ supports this response in a finite-sized system. In this work, we experimentally probe the behavior of these states.

The waveguide array that we employ (see Fig.~\ref{fig1}\textbf{b}) is fabricated using the femtosecond laser-writing technique \cite{Alex2010} such that each waveguide supports a single guided mode that can evanescently couple to its neighbors.  When light is injected into the array at the input facet, it excites the eigenmodes of the device according to its spatial overlap with them. The light, then, diffracts through the device according to the paraxial Schr\"odinger equation, $i\partial_z\psi=H(z)\psi$, where the time-evolution coordinate in the usual Schr\"odinger equation, $t$, is replaced by the distance along the axis of propagation, $z$; $\psi$ represents the excited wavefunction as a superposition of bound modes of the waveguides, and $H(z)$ is the Hamiltonian defined by the coupling in the array. Therefore, the diffraction of light through the array mimics the time-evolution of the wavefunction of a quantum mechanical particle \cite{Jason2002,Jason2003,christodoulides2003discretizing}. Consequently, time-dependent pumping as described above means adiabatically varying $\phi_i$ along the waveguide propagation axis: $\phi_i\rightarrow \phi_i(z)$~\cite{Zilberberg:2012a,Zilberberg:2015a}.     

\begin{figure}[ht]
\includegraphics[width=\columnwidth]{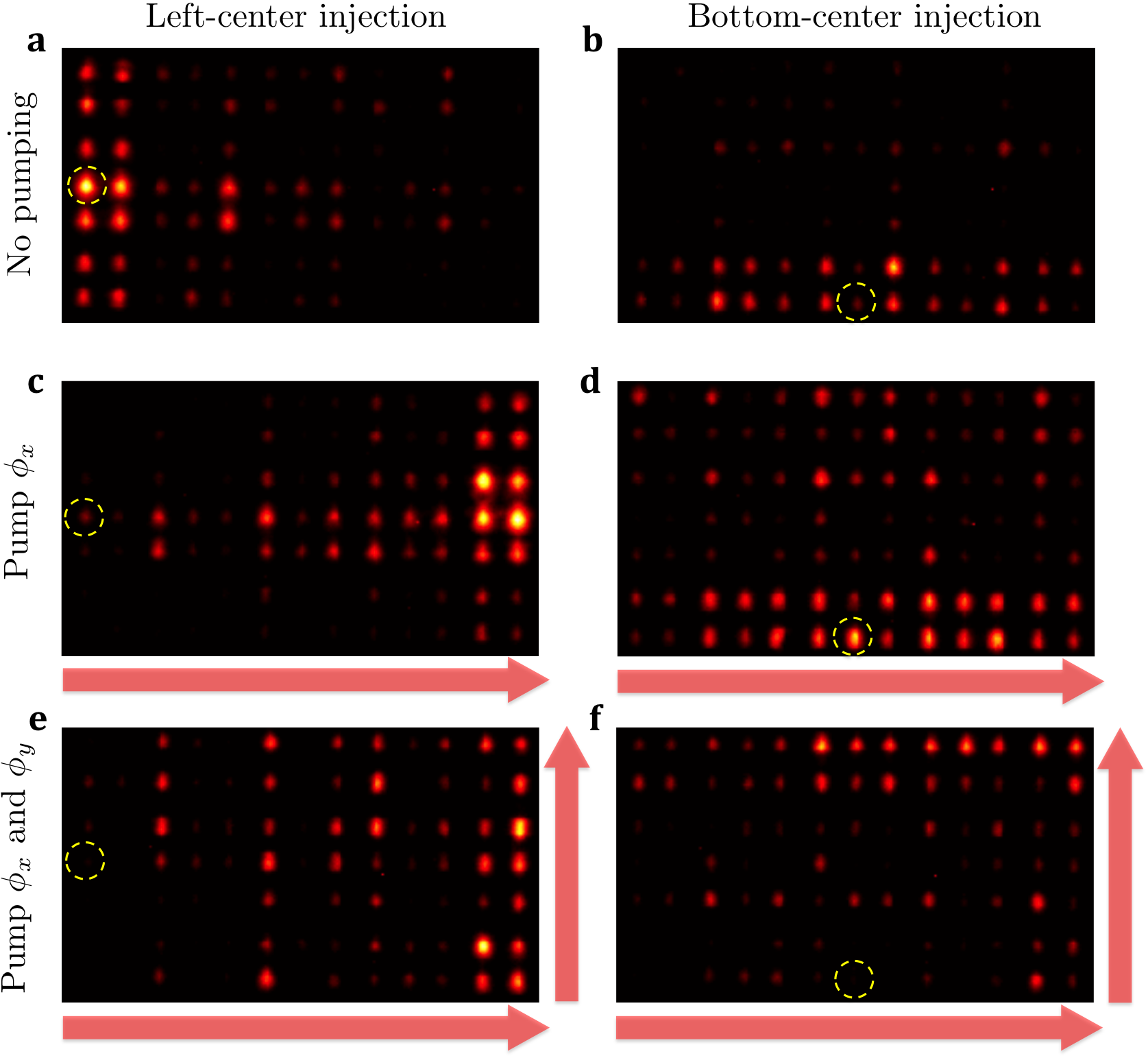}
\caption{\textbf{Images of the output facet of the waveguide arrays after $z=15cm$ of propagation showing edge-to-edge pumping.} {\textbf a},\textbf{b} A device with no pumping corresponding to a model with $\phi_x=\phi_y=0.477\pi$ [cf.~Fig.~\ref{fig1}]. Light that is injected at the center of the left or bottom edges excites the topological edge bands and spreads out along the edge, correspondingly. In {\textbf c} and \textbf{d} pumping of $\phi_x$ (from $0.477\pi$ to $2.19\pi$, while $\phi_y$ is held constant at $0.477\pi$) causes the light to be pumped from the left edge to the right, whereas no such pumping is observed when light is injected at the bottom-center.  However, when both $\phi_x$ and $\phi_y$ are simultaneously pumped (from $0.477\pi$ to $2.19\pi$), we observe that light injected at the left-center (in {\textbf e}) and bottom-center (in {\textbf f}) pump from left-to-right and bottom-to-top, respectively.  The fact that some light resides in the bulk arises from imperfect coupling to edge states as well as deviations from perfect adiabaticity. In each case, the yellow-dashed circle indicates the site of injection at the input facet, $z=0$, and the red arrows indicate the direction of photon pumping. Taken together, these results demonstrate the edge phenomena associated with the bulk response of the 2D topological pump, i.e.,  the edge states supporting the 4D symmetry predicted from the response of this dynamically generated 4D quantum Hall effect.  }\label{fig2}
\end{figure}

We experimentally demonstrate the appearance of edge modes in the structure as well as their behavior under scans of the pump parameters $\phi_x$ and $\phi_y$ as a function of $z$.  We start by studying a waveguide structure with straight waveguides, i.e., without modulation as a function of $z$. We inject light (via fiber coupling) into two different single waveguides in the array: one along the left edge and one along the bottom edge. After a diffraction length of $15cm$ the output light from the chip is collected. We see that light stays largely confined to the injected edge, i.e., it had mostly excited the topological localized edge modes (see Fig.~\ref{fig2}\textbf{a},\textbf{b}). Additionally, it spreads across the whole edge but does not show a right-angle wedge form. This implies the presence of a band of edge states (i.e., the band that crosses the gaps in Fig.~\ref{fig1}\textbf{c}), where the coupling between the two bands on the left and bottom is small. We can, therefore, conclude that the long-range coupling did not break the orientation associated with the two orthogonal 1D topological pumps embedded in the system. Having established that we can couple to the edge modes of the two perpendicular 1D pumps, we will now demonstrate their behavior under scans of the pump parameters, $\phi_i$.

We implement topological pumping by allowing the positions of the waveguides to `wiggle', corresponding a variation of $\phi_x$ and/or $\phi_y$ as a function of $z$, as depicted in Fig.~\ref{fig1}\textbf{b}. To this end, we fabricate separate arrays of waveguides that correspond to two scenarios: (1) pumping only in the $x$-direction by adiabatically changing $\phi_x$ as a function of $z$; and (2) pumping in both the $x$ and $y$ directions by adiabatically varying $\phi_x$ and $\phi_y$ simultaneously. In case (1), we see that when light is injected at the left edge, it is pumped to the right edge (see Fig.~\ref{fig2}\textbf{c}).  However, when it is injected at the bottom, it is {\it not} pumped to the top due to a lack of pumping of $\phi_y$ (see Fig.~\ref{fig2}\textbf{d}).  In case (2), we observe that the edge states pump both from left to right (Fig.~\ref{fig2}\textbf{e}) and bottom to top (Fig.~\ref{fig2}\textbf{f}).  Note that some of the edge (i.e., bulk-boundary) states that we excite lie at the same energies as bulk states in the open system geometry (see Supplementary Information); these will also lead to pumping across the sample. We note further that some occupation of bulk modes is present in all output wavefunctions - this arises from two sources: (a) some degree of overlap of the initial beam with bulk modes; and (b) deviations from adiabaticity. These results show that an electro-motive force applied in the $v$- and/or $w$-directions implies pumping from one 3D $(v,w,y)$-hyperplane to the opposite one in the $x$-direction, and/or from one 3D $(v,w,x)$-hyperplane to the opposite one in the $y$-direction as implied by the presence of a non-zero 4D $\Snd$ Chern number.  

\begin{figure}[ht]
\includegraphics[width=0.7\columnwidth]{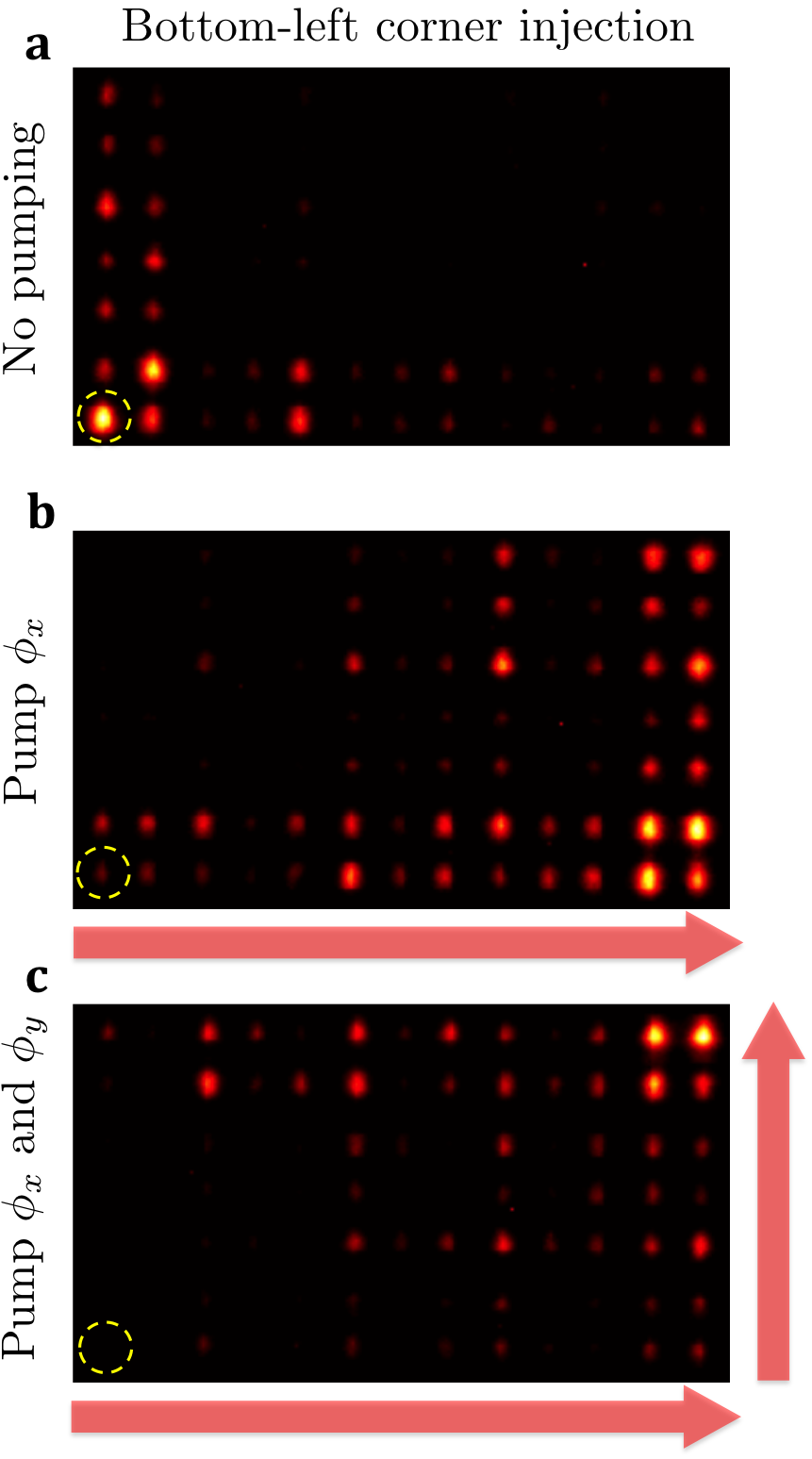}
\caption{\textbf{Images of the output facet of the waveguide arrays after $z=15cm$ of propagation showing corner-to-corner pumping.}  The devices are the same as those used in Fig.~\ref{fig2}. The yellow-dashed circle indicates the site of injection at the input facet, $z=0$, and the red arrows mark the direction of pumping. In {\textbf a}, there is no pumping, and therefore light stays confined to the corner state.  In {\textbf b}, light is largely pumped from the bottom-left corner to the bottom-right corner via $\phi_x$.  When both $\phi_x$ and $\phi_y$ are pumped (in {\textbf c}), the corner state is pumped from bottom-left to top-right.  The corner state passes through the bulk band remains localized since it is a long-lived resonance while not in the band gap, (see Supplementary Information).} \label{fig3}
\end{figure}

We now examine the pumping of states that reside at the corners of the structures described above. We inject light in the bottom-corner waveguide of each of the three devices. The corner mode can be directly excited as well as pumped along the bottom edge, in conjunction with it being the boundary mode of the 1D pump that crosses edge-to-edge, see Figs.~\ref{fig3}\textbf{a},\textbf{b}. 
Interestingly, when we scan $\phi_x$ and $\phi_y$ simultaneously, the corner mode is pumped largely all the way to the top-right corner of the sample, see Fig.~\ref{fig3}\textbf{c}. In both cases, the corner mode becomes degenerate with bulk states and should hybridize with them (see Supplementary Information). Nevertheless, as each constituent 1D pump is characterized by its own $\Fst$ Chern number (in dispersive indirect gaps), the corner modes manifest as the joint product of the protected topology at the boundary of the 2D pump.  This means that the corner modes only weakly hybridize with the bulk, since in the limit where the system can be separable into a tensor product, the corner states are fully bound. Hence, we interpret the in-bulk-band corner modes as long-lived resonances that have lifetimes much greater than the time over which pumping is carried out, i.e., the $15cm$ of propagation. In the 4D picture, the 2D corner modes correspond to two-dimensional `hyper-edges' that are extended in the $(v,w)$-plane.  Pumping of the corner modes therefore corresponds to the pumping of one hyper-edge to the one diagonally across from it in the $(x,y)$-direction. Note that the in-gap corner modes are unique in the sense that they are topologically protected modes that are two dimensions lower than the physical dimension of the system (conventional protected modes are one dimension lower). Recently, the appearance and demonstration of such modes has been reported in inversion-symmetry protected 2D systems  \cite{benalcazar2016quantized,noh2016topological}.

In conclusion, we have observed topological edge pumping associated with the 4D quantum Hall effect in a 2D photonic system using synthetic dimensions, implying the presence of a non-zero $\Snd$ Chern number characterizing the system.  The use of edge/surface physics provides an independent observation of the physics implied by the system's $\Snd$ Chern number, as compared with the measurement of quantized nonlinear bulk response in a similar model studied in a concurrent work using cold atoms [M.~Lohse \textit{et al.}, submitted].  The realization of 4D quantum Hall physics opens up the possibility of many new physical effects.  Natural questions include: can arbitrarily high spatial dimensionality be realized?  Can interactions lead to 4D fractional Hall physics using synthetic dimensions?  Since photonic systems naturally allow for non-Hermitian Hamiltonians (arising from gain and loss), what is the interplay between non-Hermiticity and topological gaps associated with non-zero $\Snd$ Chern number?  Are there other physical quantities that are quantized in four dimensions that can be directly measured using synthetic dimensions?  We expect that with new experimental access to 4D quantum Hall physics, many other directions will emerge.


%

\vspace{5mm}
\textbf{Acknowledgments} We thank H.~M.~Price and M.~Lohse for feedback on the manuscript. O.Z. thanks the Swiss National Science Foundation for financial support.  M.C.R. acknowledges the National Science Foundation under award number ECCS-1509546 and the Alfred P. Sloan Foundation under fellowship number FG-2016-6418. K.P.C. acknowledges the National Science Foundation under award numbers ECCS-1509199 and DMS-1620218. 
  
\vspace{5mm}
\begin{center}
\textbf{\large Methods}
\end{center}
\textbf{Experimental specifications.} The experiments were conducted using arrays of evanescently coupled
waveguides fabricated in borosilicate glass using the femtosecond laser writing technology \cite{Alex2010}. The waveguides
are all identical in both refractive index and dimension, while the
inter-waveguide separation was modulated in order to realize the off-diagonal 2D 
model described in the text [i.e., Eq.~\eqref{Eq:H2D}].  In all cases, we observe the output image (after $15cm$ of propagation) over a range of wavelengths (i.e., between 1510nm-1590nm) in increments of 5nm, and then average the output intensities over all wavelengths, see Figs.~\ref{fig2} and \ref{fig3}. Since the band gap remains open over this range, we expect the pumping behavior to remain the same.  We perform the averaging over wavelength in order to minimize sensitive interference effects due to fabrication imperfections.  


\textbf{Model implementation with waveguide arrays.}
The diffraction of paraxial light through the structures described in the text is governed by the paraxial Schr\"odinger equation
\begin{equation}\label{paraxial_schrodinger}
i \partial_z \psi= -\frac{1}{2k_0} \nabla^2 \psi - \frac{k_0 \Delta n}{n_0} \psi \nonumber
\end{equation}
where the wavefunction $\psi(x,y,z)$ corresponds to the electric field envelope $\mathbf{E}(x,y,z) = \psi(x,y,z) \exp(i k_0 z - i\omega t) \hat{\mathbf{E}}_0$. In the above, $\nabla^2  = \partial_x^2 + \partial_y^2$ is the transverse Laplacian, $\Delta n(x,y,z)$ is the change in refractive index relative to the background index $n_0$, and $k_0 = 2\pi n_0 / \lambda$ is the wavenumber in the background medium.  For an array of single mode, weakly coupled waveguides, the evolution generated by the paraxial Schr\"odinger equation can be described using tight-binding theory, where light hops between the bound modes of adjacent waveguides. The hopping amplitude $t$ associated with a given waveguide separation can be obtained by numerically computing the two lowest eigenvalues  $E_1, E_2$ of the full equation for a system consisting of two waveguides. The hopping amplitude is then given by $t = (E_2 - E_1)/2$.

\begin{figure}[ht]
\includegraphics[width=0.5\textwidth]{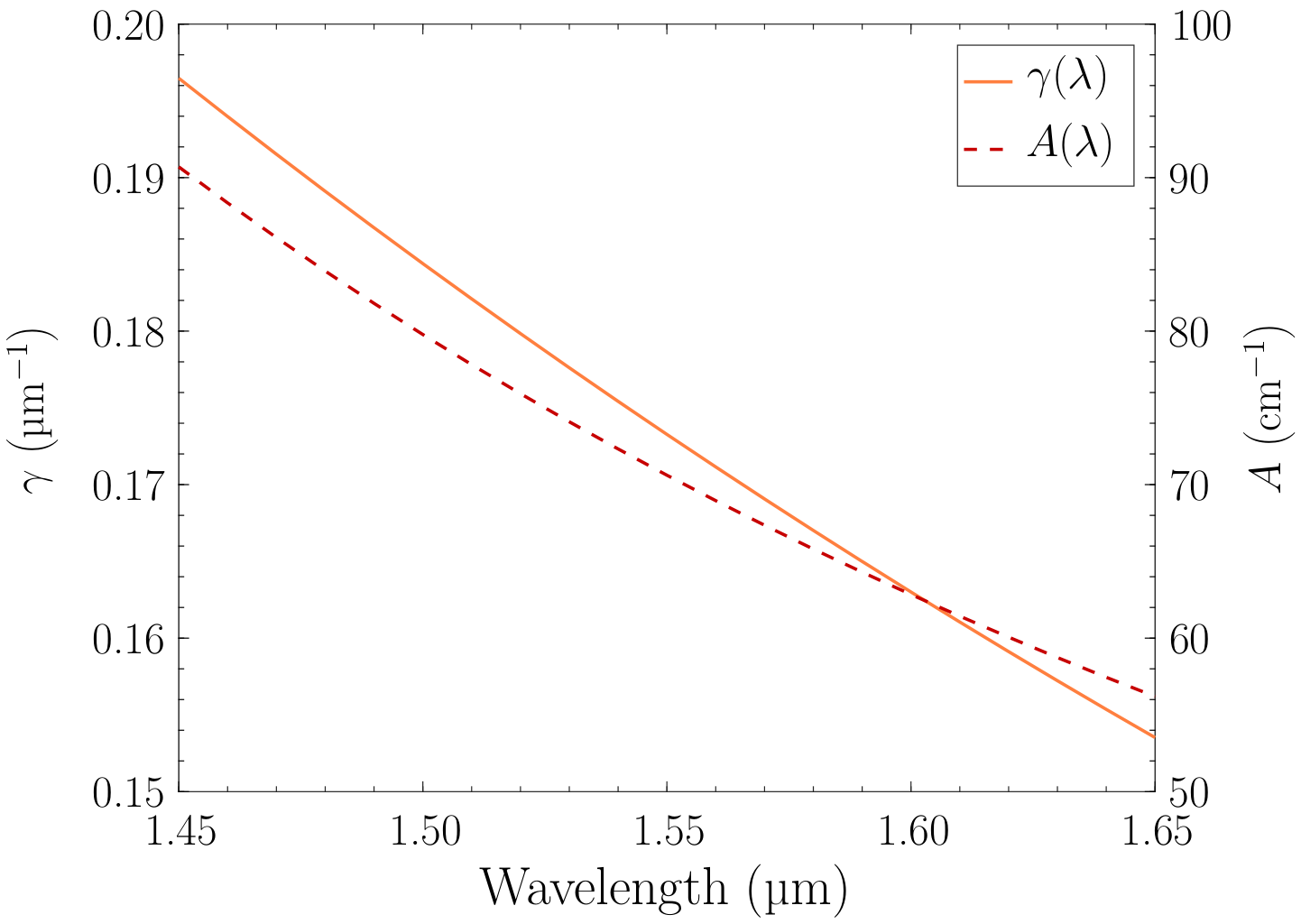}
\caption{The overall scale $A$ (dashed red line) and exponential decay prefactor $\gamma$ (solid orange line) describing the inter-waveguide coupling versus their separation $t(s) = A \exp(- \gamma s)$, where $s$ is the waveguide separation. The parameters were obtained using a thorough calibration procedure (see Methods) and are plotted as a function of wavelength.}\label{figSupp2}
\end{figure}

To perform this computation for our waveguides, we used a best-fit Gaussian model for the waveguide refractive index variation:  $\Delta n(x,y) = \delta n \exp(-x^2/\sigma_x^2 - y^2/\sigma_y^2)$ with $\delta n = 2.8\times10^{-3} , \sigma_x = \SI{3.5}{\micro \metre}$, and $\sigma_y = \SI{5.35}{\micro\metre}$.  These parameters were obtained through calibration over a set of one-dimensional test arrays. Using this profile and a background index of $n_0=1.473$, we obtained a model of the form $t(s) = A \exp(- \gamma s)$ governing the dependence of the hopping amplitudes on the waveguide separation, $s$. Here  $A=A(\lambda)$ and $\gamma =\gamma(\lambda)$ are wavelength dependent parameters plotted in Fig.~\ref{figSupp2}. We then used this model to solve for the waveguide spacings required to implement the modulated hopping amplitudes defined by the Hamiltonian in Eq.~\eqref{Eq:H2D}.

\newpage
\cleardoublepage
\setcounter{figure}{0}
\renewcommand{\figurename}{Supplementary Information Figure}

\onecolumngrid
\begin{center}
\textbf{\normalsize Supplemental Information for:}\\
\vspace{3mm}
\textbf{\large Photonic topological pumping through the edges of a dynamical four-dimensional quantum Hall system}
\vspace{4mm}

{ Oded Zilberberg,$^{1}$ Sheng Huang,$^{2}$ Jonathan Guglielmon,$^{3}$ Mohan Wang,$^{2}$\\
Kevin Chen,$^{2}$ Yaacov E. Kraus,$^{4,\dagger}$ and Mikael C. Rechtsman,$^{3}$}\\
\vspace{1mm}
\textit{\small $^{1}$Institute for Theoretical Physics, ETH Zurich, 8093 Z{\"u}rich, Switzerland\\
$^{2}$Department of Electrical and Computer Engineering, University of Pittsburgh, Pittsburgh, Pennsylvania 15261, USA\\
$^{3}$Department of Physics, The Pennsylvania State University, University Park, Pennsylvania 16802, USA\\
$^{4}$Department of Physics, Holon Institute of Technology, Holon 5810201, Israel}

\vspace{5mm}
\end{center}
\setcounter{equation}{0}
\setcounter{figure}{0}
\setcounter{table}{0}
\setcounter{page}{1}
\makeatletter
\renewcommand{\theequation}{S\arabic{equation}}
\renewcommand{\bibnumfmt}[1]{[S#1]}
\renewcommand{\citenumfont}[1]{S#1}

\twocolumngrid

\begin{center}
\textbf{\large The direct sum model}
\vspace{3mm}
\end{center}

The direct sum model [Eq.~\eqref{Eq:H2D} in the main text] decomposes along the $x$- and $y$-directions into a sum of two independent off-diagonal Harper models~\cite{supp_Harper:1955,supp_Zilberberg:2012a,supp_Zilberberg:2013b}
\begin{equation} \label{eqn_tensor_Hamiltonian}
H(\phi_x,\phi_y) = H_x(\phi_x) + H_y(\phi_y)
\end{equation}
where $H_i(\phi_i)  = \tilde{t}_i + \lambda_i\cos(2\pi b_i\, i + \phi_i)$ with $i \in \{x,y\}$. Each $H_i(\phi_i)$ is a one-parameter family of 1D Hamiltonians, i.e., a 1D topological pump. Treating the parameter $\phi_i$ as a Bloch momentum associated with an additional spatial dimension $\tilde{i}\in \{v,w\}$, we can perform a dimensional extension of this model and obtain a model describing the 2D integer quantum Hall system on a lattice with anisotropic hopping~\cite{supp_Zilberberg:2012b,supp_Hofstadter,supp_TKNN}.

For $b=1/3$, the spectrum of the [1D pump $\leftrightarrow$ 2D QH] system consists of three bands as shown in Supplementary Information Fig.~\ref{figSupp}\textbf{a}. Each band, $n$, has an associated non-zero $\Fst$ Chern number 
\begin{align} \label{Eq:Chern}
\nu_n & =\frac{1}{2\pi i}\int_{0}^{2\pi}d\phi_i dk_i\, C_n(\phi_i,k_i)\,,
\end{align}
 which is an integral over the Berry curvature (also known as the Chern density) of the filled $n^{\rm th}$ band
\begin{align}
C_n(\phi_i,k_i) & =\textrm{Tr}P_n\left[\frac{\partial P_n}{\partial\phi_i},\frac{\partial P_n}{\partial k_i}\right]\,,\label{Eq:Chern-1}
\end{align}
where we have defined the spectral projector $P_n$ onto all states in the $n^{\rm th}$ band. The $\Fst$ Chern number of the bands manifests through the quantization of the Hall conductance in response to an applied in-plane electrical field, e.g., in our case $I_{x}=(e^2/h) E_v\sum_n \nu_n $, where $I_x$ denotes the current density along $x$-direction, $E_v$ is an electric field along the $v$-direction, and the sum is over all filled bands. This quantized bulk response has corresponding edge states, i.e., gapless boundary states appear in a finite sample (as many as the sum of Chern numbers below a given gap) and carry the transverse quantized conductance~\cite{supp_Hatsugai}. 

As discussed in the main text, the eigenstates of the full Hamiltonian [Eq.~\eqref{Eq:H2D} in the main text and Eq.~\eqref{eqn_tensor_Hamiltonian}] are tensor products of the eigenstates of the two independent Harper models $|\psi_{mn}\rangle = |\psi_{m}\rangle \otimes |\psi_n\rangle$, where $m$ enumerates the states in the $x-v$ plane and $n$ those in the $y-w$ plane. Their associated energies are $E_{mn} = E_m + E_n$ so that each pair of bands from the decoupled models yield a band of the [2D pump $\leftrightarrow$ 4D QH] model. Therefore, in a finite system, as each constituent of the direct sum has bulk and boundary modes, the tensor product eigenstates can be categorized as bulk-bulk, bulk-boundary, and boundary-boundary. A color-coded illustration of the resulting band structure is shown in Supplementary Information Fig.~\ref{figSupp}\textbf{b}. 

It is important to note that the resulting Minkowski sum spectrum is not always gapped, i.e., depending on the amplitudes $\tilde{t}_i$ and $\lambda_i$, the joint spectrum may not be gapped. Consequently, if the gaps are closed, we can no longer discuss the topology of the combined spectrum as any small perturbation will mix the states from the different bands. When the spectral gaps are open, we note that the bulk-boundary and boundary-boundary modes lie for some $\phi_i$ at energies within the gaps and for others at energies in the bulk bands. Therefore, the boundary-boundary (i.e., 2D corner) modes that overlap with the bulk are generally expected to become finite lifetime resonances upon the introduction of higher-neighbor couplings that destroy the tensor product structure of the eigenstates.  Nonetheless, the in-gap bulk-boundary and boundary-boundary modes are protected for arbitrary perturbations that do not close the gap and are the surface states associated with a non-zero $\Snd$ Chern number. 

\begin{figure*}
\centering
\includegraphics[width=0.9\textwidth]{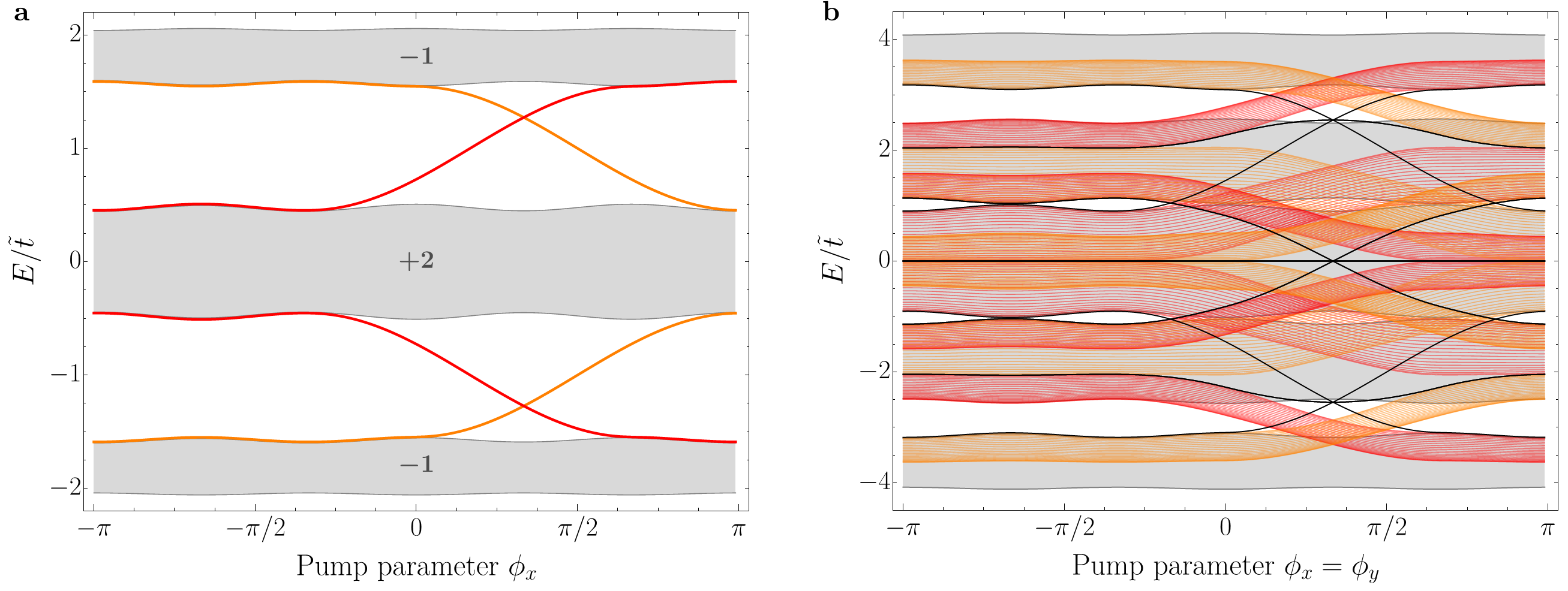}
\caption{\textbf{Nearest neighbor band structure obtained from two decoupled models [see Eq.~\eqref{eqn_tensor_Hamiltonian}].} \textbf{a,} Finite sample band structure for a single Harper model aligned along the $x$-direction. Boundary modes highlighted in orange (red) are localized on the left (right) end of the 1D sample. Also shown is the first Chern number associated with each bulk band. \textbf{b,} Band structure taken along the path $\phi_x=\phi_y$ for a system that decomposes into two independent Harper models.  Each band in the figure on the right is obtained by summing together a pair of bands from the figure on the left. The resulting bands can be classified by the types of states appearing in the sum: bulk-bulk (2D bulk), bulk-boundary (2D edge), and boundary-boundary (2D corner). These types are respectively colored gray, red/orange, and black. Note that as a function of $\phi$ the edge modes form ``dispersive'' bands that thread through the 2D bulk gaps. We identify that the edge bands have indirect gaps between them, in conjunction with the fact that on the edge of the 2D model a 1D pump exists that has a 1D spectrum as appearing in \textbf{a}. The corner modes are threading between the edge bands and, therefore, are forced to cross 2D bulk bands along their $\phi$ trajectory. }\label{figSupp}
\end{figure*}

\vspace{3mm}
\begin{center}
\textbf{\large Second Chern number bulk response and its corresponding edge phenomena}
\vspace{3mm}
\end{center}

Let us consider an energy $E_j$ that resides in the $j^{\rm th}$ gap of the [2D pump $\leftrightarrow$ 4D QH] system, see Supplementary Information Fig.~\ref{figSupp}\textbf{b}. The second Chern number $\mathcal{V}_j$ associated with this gap is given by 	
\begin{equation}
\mathcal{V}_j = -\frac{1}{8\pi^2} \int d^4k\, \epsilon_{\mu\nu\rho\sigma} \text{tr}\left(P_j\, \frac{\partial P_j}{\partial k_\mu}\frac{\partial P_j}{\partial k_\nu}P_j\frac{\partial P_j}{\partial k_\rho}\frac{\partial P_j}{\partial k_\sigma}\right)
\end{equation}
where $P_j(k_\mu)$ is the projector onto the subspace spanned by the eigenstates at Bloch momentum $k_\mu=(\phi_x,\phi_y,k_x,k_y)$ with energies below the gap. Using the decomposition of $H$ discussed above, $\mathcal{V}_j$ can be written in terms of the first Chern numbers $\nu_n$ of the Harper models as: \cite{Zilberberg:2012a}
\begin{equation}
\mathcal{V}_j = \sum_{{\text{band pairs $m,n$ with total energy}} < E_j} \nu_n \nu_m\,.
\end{equation}
Combining this result with the $\Fst$ Chern numbers shown in Supplementary Information Fig.~\ref{figSupp}\textbf{a}, the $\Snd$ Chern numbers associated with the lower and upper gaps of the [2D pump $\leftrightarrow$ 4D QH] Hamiltonian can be seen to be $\mathcal{V} = +1,-1$, respectively. While the Hamiltonian governing our photonic system does not decompose in the way discussed above due to the presence of higher neighbor couplings, the upper and lower gaps remain open (see Figure \ref{fig1} in the main text) and, as a result, the associated second Chern numbers remain unchanged.

The $\Snd$ Chern number has an associated quantized nonlinear bulk response $I_\alpha = \mathcal{V}_j \frac{e^2}{h\Phi_0} \epsilon_{\alpha\beta\gamma\delta} B_{\beta\gamma} E_\delta$~\cite{QiZhang}, where $I_\alpha$ denotes the current density along the $\alpha$ direction, $\Phi_0$ is the flux quantum, $E_\delta$ is an electric field along the $\delta$ direction, and $B_{\beta\gamma}$ is a magnetic field in the
$\beta\gamma$ plane. This response has a distinct 4D symmetry to it, namely, the topology implies that all of the various responses associated with the $\Snd$ Chern number must occur with the exact same quantization. In turn, in a finite-sized system, boundary states must appear and support these quantized bulk responses. 

We can split the $\Snd$ Chern number bulk responses of the direct sum model into two types: density-type responses and Lorentz-type responses~\cite{supp_Zilberberg:2015b,supp_Zilberberg:2016a,supp_Zilberberg:2016b}. The former occurs when the magnetic field perturbation is applied in a plane coinciding with that of one of the intrinsic magnetic fields that generate the topological gaps. The perturbation, thus, changes the density of one of the constituting direct sum models according to the Streda formula~\cite{supp_Streda82}. This means that the density of the bulk-boundary in-gap states would change and carry the additional quantized response associated with the $\Snd$ Chern number. As this response occurs for any infinitesimal magnetic field perturbation and is additive to the $\Fst$ Chern number response that each constituent model should contribute in response to an applied electric field perturbation, we therefore see in our experiments edge-to-edge transport carried by these states that support density-type bulk responses. The latter occurs when the magnetic field perturbation is in a plane orthogonal to the intrinsic magnetic fields. Here, the two direct sum models are coupled through the magnetic field perturbation leading to mutual scanning of $\phi_i$. This response is carried by the corner states.


%


\end{document}